\documentclass[a4paper]{jpconf}
\usepackage{amsmath,amsfonts,amssymb,epsfig,color}

\usepackage{graphicx}
\def\be{\begin{equation}}
\def\ee{\end{equation}}
\def\ba{\begin{eqnarray}}
\def\ea{\end{eqnarray}}

\begin{document}
\title{Description of heavy deformed nuclei within the pseudo-SU(3) shell model}

\author{Gabriela Popa$^1$  and Ana Georgieva$^2$}

\address{$^1$Ohio University Zanesville, Zanesville, OH 44906, USA}
\address{$^2$Bulgarian Academy of Science, Sofia, Bulgaria}

\ead{popag@ohio.edu}

\begin{abstract}
We present a review of the pseudo-SU(3) shell model and its application to heavy deformed nuclei.
The model have been applied to describe the low energy spectra, B(E2) and B(M1) values. A systematic study
of each part of the interaction within the Hamiltonian was carried out. The study leads us to a consistent
method of choosing the parameters in the model.
A systematic application of the model for a sequence of rare earth nuclei demonstrates that an
overarching symmetry can be used to predict the onset of deformation as manifested through
low-lying collective bands.The scheme utilizes an overarching sp(4,R) algebraic framework.
\end{abstract}

\section{Introduction}

Over the past  years many calculations were carried out by using the SU(3) shell model and
the pseudo-SU(3) symmetry. Ongoing improvements and innovations in experimental techniques
(4$\pi$ detectors, radioactive beams, etc.) anticipate the identification of additional new
phenomena in the near future, and more information about the existing ones. However, it is commonly
accepted that the nuclear shell model should be able to address most of these issues and provide
answers. The problem is that most shell-model theories are limited by the large dimensions of
the required model spaces, which increases unmanageably
with the number of nucleons $A$.  Furthermore,
more microscopic calculations are needed  exactly in the region of heavy nuclei,
in the mass range A $\geq$ 100.

Having this problem, it is essential to take full advantage of symmetries, those that are exact,
as well as those which are only
partially fulfilled.The selection rules that are associated with
such symmetries generate, respectively, weakly coupled and disconnected subspaces of the full space
and this allows for a significant reduction in the dimensionality of the model space. 
The symmetry
used here is the pseudo-spin symmetry. 

A shell model theory for heavy nuclei requires a severe truncation of the model space. To reproduce
the essential physics found in the low-energy states of a large space in a smaller one, we have to
select the basis states relative to those parts of the interaction that dominate the low-energy
structure. Nuclear physics supports the view that the nuclear effective interaction appropriate to
low-energy excitations must have a strong correlation with the pairing and quadrupole-quadrupole
interactions. The $Q \cdot Q$ interaction, which dominates for near mid-shell nuclei, is known to
introduce deformation and led to the introduction of the SU(3) shell model. This led also to
a very natural way to truncate large model spaces, namely to consider few basis states  
in correspondence with the eigenvalue of $Q \cdot Q$ operator.

Based on the early contribution of Bohr and Mottelson\cite{BM53}, Elliott\cite{Elliott58} 
developed an SU(3) symmetry truncation for light $ds$-shell nuclei.
Raju, Draayer, and Hecht\cite{RDH73} developed a pseudo-SU(3) scheme for heavy nuclei.
Ground and gamma band rotational structure of deformed nuclei were described in 1984 
by Draayer and Weeks\cite{Draayer1984}, in the framework of the
pseudo-SU(3) model by using only one SU(3) irreducible representation (irrep),  the one 
 corresponding to the highest
deformation, and considering the abnormal parity spaces to seniority zero configurations. The fits to the
available data were very accurate ($|E_{exp}-E_{thy} \leq 25 keV $).  In this small space the ground and 
the gamma (K=2) bands were very well described,  as well as  the inter-band and intra-band B(E2) strengths.
 The model used the mathematical and computational advances made by Draayer and Williams\cite{Draayer1968},
Akiyama  and Draayer\cite{Akiyama1973},
Draayer, Weeks, and Hecht\cite{DraWH82},
  and Draayer and Rosensteel\cite{DraayerR1982}.

Using  mathematical and computational developments to the pseudo-SU(3) model, Casta${\tilde n}$os, Draayer, and Leschber   improved it further to accommodate an increased model space, taking into 
consideration several protons and neutrons SU(3) irreps, as well as the total coupled representation in 1987\cite{CasDraLes1987npa, CasDraLes1987ann}.  In this version, many bands can be calculated and as well the B(E2) and B(M1) transitions values within and between bands. Strong B(M1) transitions were predicted first in various Gadolinium, Erbium, Uranium, and
Plutonium isotopes using the pseudo-SU(3) model in 1987\cite{CasDraLes1987npa}.
Naqvi and Draayer introduced a shell-model operator for K-band splitting in \cite{NaqD90}. 
The $K^2$ operator is used to resolve multiple occurrences of L-values in SU(3) representations. 
In 1994, Bahri and Draayer completed an improved code for faster calculations\cite{BahD94}. 
Later the effects of paring interaction on the K-band mixing were studied\cite{TrBahDra95, TroBahEscDra1995} 
by Trotelnier, Bahri, Escher and Draayer
within the framework of the pseudo-SU(3) model.

Considering such 
larger spaces with an SU(3) Hamiltonian plus proton and neutron single particle energies,  and pairing
interactions, an extended version of the model was applied\cite{BHD00, 
PHD00}   to describe the low-lying 
energy structure  and electromagnetic transitions probabilities of even-even nuclei in the  rare-earth region . 

The purpose of this work is to show that by using a realistic Hamiltonian based on proton and neutron
single-particle energies,  $Q \cdot Q$, and pairing interactions we can find systematically a 
set of parameters\cite{PHD00}, in the chosen model space, to obtain a  good agreement with the experiment in the even-even heavy nuclei. We are also
studying the effect of each term in the Hamiltonian plus the rotor terms, namely $J^2$, $K^2$, and
$Q_3$, on the energy spectra. The parameters of the rotor terms are fixed as the
result of this study.

The role of these parameters is further revealed in a 
systematic study of the behavior of first excited $K^\pi=0_2^+$, and
$K^\pi=2^+$ bands in deformed even-even nuclei  of the rare-earth region.  The nuclei  are
ordered in F-spin multiplets of a Sp(4,R) classification scheme
\cite{Drenska95}. The energy levels of the ground band (g.b.)
$J=2_1^+$, first excited $J^\pi=0^+_{K^\pi=0^+_2}$,
$J^\pi=2^+_{K^\pi=2^+_1}$ states  of nuclei that belong to the
$F_0 =0 = 1/2(N^\pi -N^\nu)$ multiplet (where $N^\pi$ and $N^\nu$
are the number of valence proton and neutron pairs) are plotted in
Fig. \ref{fig:1}. The energies of the same set of levels for
nuclei belonging to the $F_0=1$ multiplet are plotted in Fig.
\ref{fig:2}. Nuclei which belong to the  $F_0=0$  multiplet have equal numbers of valence
proton and neutron pairs, and those which belong to the $F_0=1$ multiplet vary by two pairs
of protons. With such an investigation\cite{Popa04}, we provide a  microscopic interpretation of
the complex and varying behavior of these collective excitations, by  applying the
algebraic shell model with pseudo SU(3) symmetry.

In the next section, the structure of the basis states is defined. 
An example is given for $^{156}$Gd nucleus.
The Hamiltonian is described in detail in section 3. 
The results of the analysis are explained in section 4. 
Also further applications of the model are presented.

\newpage
\section{Basis states} Building the many-body basis of the pseudo-SU(3) model relies on
the proton-neutron degrees of freedom, pseudo-spin symmetry and group theoretical methods. The
construction of the basis states starts with the  deformed Nilsson single-particle levels for
protons and neutrons, which are filled from the lowest energy  upward,
for a specific deformation parameter
$\beta$ for the corresponding nucleus. The completely filled shells are frozen and the remaining
levels are taken into consideration.  For heavy nuclei, the valence protons and neutrons fill different  major shells.
For a given nucleus there are two open shells (one for protons, one for neutron), each of them comprised of a set of normal and unique parity levels. The normal parity space is partitioned into pseudo SU(3) irreps, and the unique parity space is spanned by configurations of identical particles in a single j shell. It was shown that the seniority coupled scheme is appropriate for a description of the unique parity configuration\cite{DraWH82}. Thus the low-energy structure of the normal parity part of the space will be dominated by a few irreps of pseudo SU(3). In the abnormal(unique) parity parts of the proton and neutron shells was shown that the configuration with low seniority are the most important ones, since those with high seniority are not favored (the pairing gap is large compared to the spacing of low-lying  rotational bands.)

In this way, the  occupation numbers of the normal ($N$) and unique ($U$)(also called intruder)
parity states are determined for protons ($\pi$) and neutrons ($\nu$), $n^N_\pi$, $n^U_\pi$, $n^N_\nu$, $n^U_\nu$. 
For example, for
  $^{156}Gd$, $n^N_\pi = 8$
in $\tilde \eta_\pi = 3$, and $n^N_\nu =6$ in $\tilde \eta_\pi =
4$ with degeneracies $\Omega_\pi = 20$ and $\Omega_\nu = 30$.
 Here, $\tilde \eta_\pi[\nu]$, represent the pseudo associated numbers.
 For 8
protons in $\tilde \eta_\pi = 3$ and 6 neutrons in $\tilde \eta_\nu = 4$ there are 32
possible SU(3) irreps in the proton space and 48 possible irreps in the neutron space, for
proton and neutron pseudo-spin equal to zero. The corresponding irreps are ordered in descending $C_2$
value and the first 6 proton and 5 neutron irreps are chosen, that are given in Table \ref{t:pnirrepsGd156}.
\begin{table*}[h]
\begin{center}
\caption{ The SU(3) irreps for protons and neutrons corresponding to the largest
$C_2$ in $^{156}Gd$.}
\begin{tabular}{cc|cc}
\br
$(\lambda_{\pi},\mu_{\pi})$ & $C_2$ &
$(\lambda_{\nu}, \mu_{\nu})$ & $C_2$ \\
\mr
 (10,4) & 198  & (18,0)    & 378\\
 (12,0) & 180  & (15,3)    & 333\\
 (0,12) & 180  & (12,6)    & 306\\
 (8,5)  & 168  & (13,4)    & 288\\
 (5,8)  & 168  & (14,2)    & 276\\
 (9,3)  & 153  &           &    \\
 \br
 \vspace{-10mm}
\label{t:pnirrepsGd156}
\end{tabular}
\end{center}
\end{table*}
The direct products of these proton and neutron irreps give a large set of possible strong coupled
irreps from which we have selected the first 30 with the largest $C_2$ values. These irreps are
given in Table \ref{t:irrepsGd156}. The many-particle states of $n_{\alpha}$ nucleons in a shell ($\alpha$ = $\pi$ or $\nu$)
belong to the totally antisymmetric irreps   $\{ 1^{n_\alpha} \}$ of the unitary group of $\Omega_{\alpha}^N$ =
($\eta_{\alpha} +1$)($\eta_{\alpha} +2$) dimension. A complete classification of the basis states uses the
quantum numbers following from subgroups of the unitary group.
\ba
\{ 1^{n^{N}_\alpha} \} ~~~~~~~~ \{ \tilde{f}_\alpha \} ~~~~~~\{ f_\alpha
\} ~\gamma_\alpha ~ (\lambda_\alpha , \mu_\alpha ) ~~~ \tilde{S}_\alpha
~~~~~~~K_\alpha  \tilde{L}_\alpha  ~~~~~~~~~~~~~~~~~~ J^N_\alpha ~~~~ \nonumber \\
U(\Omega^N_\alpha ) \supset U(\Omega^N_\alpha / 2 ) \times U(2) \supset SU(3) \times SU(2) \supset
SO(3) \times SU(2) \supset SU_J(2),
\label{eq:chains} \ea
\noindent Above each group  are given the quantum numbers that characterize its irreps with $K_{\alpha}$ and
$\gamma_{\alpha}$ the multiplicity label of the indicated reduction. The eigenstates $|\psi_{\alpha}>$ of the
Hamiltonian~(\ref{eq:basist})   are written as linear combination of the strong-coupled basis:

\ba
|\psi_{\alpha}> = \sum_i C^{\alpha}_i |\phi_i>\label{eq:basist}
\ea
\ba
|\phi_i> = |\{n_\pi[f_\pi](\lambda_\pi,\mu_\pi),S_\pi;
n_\nu[f_\nu](\lambda_\nu,\mu_\nu),S_\nu
\}\rho(\lambda,\mu)KL~S;JM>,\label{eq:basiss}
\ea
\noindent with two multiplicity labels: $\rho$ that counts the number of times the irrep ($\lambda,\mu$) occurs
in the direct product $(\lambda_\pi,\mu_\pi)(\lambda_\nu,\mu_\nu)$ and K classifies the different occurrences of
the orbital angular momentum L in $(\lambda,\mu)$.
\begin{table*}[h]
\begin{center}
\caption{The first 30 pseudo SU(3) irreps used in the description of $^{156}Gd$ bands.}
\begin{tabular}{cc|cccccc}
\br
$(\lambda_{\pi},\mu_{\pi})$ &
$(\lambda_{\nu}, \mu_{\nu})$ &
\multicolumn{6}{c}{total $(\lambda, \mu)$ }\\
\mr
 (10,4) & (18,0) & (28,4) & (26,5) & (27,3) &(24,6) & & \\
 (10,4) & (15,3) & (25,7) & (26,5) & (27,3) &(28,1) & (23,8)& (24,6) \\
 (10,4) & (12,6) & (22,10)& (23,8) & (24,6) &(20,11)\\
 (10,4) & (13,4) & (23,8) & (24,6) & \\
 (10,4) & (14,2) & (24,6)\\
 (12,0) & (18,0) & (30,0) & (28,1) \\
 (12,0) & (15,3) & (27,3)\\
 (12,0) & (12,6) & (24,6)\\
 (8,5)  & (18,0) & (26,5) & (24,6)\\
 (8,5)  & (15,3) & (23,8) & (24,6)\\
 (5,8)  & (18,0) & (23,8)\\
 \br
 \vspace{-10mm}
\end{tabular}\label{t:irrepsGd156}
\end{center}
\end{table*}

\section{Model Interaction}
We studied in detail the effect of
each term of the Hamiltonian on the energy spectra. The relative strength of each term in the interaction affects not only the energy spectra, but also the strength of the B(E2) and B(M1) transitions among those states.
The Hamiltonian consists of quadrupole-quadrupole ($Q \cdot Q$), proton ($\pi$) and neutron ($\nu$) single particle energies ($H^{\pi}_{sp} +H^{\nu}_{sp}$),   proton and neutron pairing
 ( $H^{\pi}_{P}$ and  $ H^{\nu}_{P}$),  as well as the rotor part ($H_{ROT}$).

\begin{equation} H =   \chi Q \cdot Q +H^{\pi}_{sp} +H^{\nu}_{sp}   -G_\pi H^{\pi}_{P} -G_\nu H^{\nu}_{P}
+ H_{ROT}\label{eq:totalhm}
\end{equation}
where
\be
H_{ROT} = aJ^2 + b K^2_J + c_3 C_3 + a_{sym}C_2  \label{eq:vrot}
\ee 
$C_2$  and $C_3 $ terms are the second and third order Casimir invariants of SU(3), which are related to the
deformation of the nucleus.  $J^2$ is the total angular momentum operator, and $K^2_J$ is the K-band splitting operator(resolving the multiple occurrences of the same angular momentum within an SU(3) irrep). The single particle energies are calculated in the standard way:

\begin{equation}
H^{\sigma}_{sp} = H_{0} + c \hat l \hat s + d \hat l^2 = H_{OSC} - \hbar \bar \omega_0 \kappa ( 2 \hat l.\hat s
+\mu
\hat l^2) \label{eq:hSP}
\ee
\noindent where $\sigma = \pi$ or $\nu$, $H_0$ is the spherical oscillator for which SU(3) is an exact symmetry, and the constants values
are taken from \cite {Ring79}. For $^{156}Gd$ the
$\hbar\omega_0$ = 41 MeV $\times A^{-1/3} = 7.616 MeV$ is the oscillator constant for the equivalent spherical nucleus and the
constants $\kappa$ and $\mu$ have the values \cite{Ring79}: $
\kappa_\pi = 0.06370, \kappa_\nu =0.60, \mu_\pi = 0.06370, \mu_\nu =0.42
$

\section{Results and conclusions}
\label{sec:3}

We studied the contribution of each term in the total Hamiltonian eq. (\ref{eq:totalhm}) by varying the strength of each interaction and plotting the energy of the first few states of angular momentum $K=0,2,4,6,8,$ and $1$. In Figure \ref{gd156envGp} we present the energy values of few states of total angular momentum $J=0,2$, and $4$ versus the strength of the proton and neutron paring. In the calculation, the strength of the proton and neutron paring is set the same and varied. The paring interaction is important to get the correct relative value of the states in the $K=0_2^+$ band, versus the ground band. It is also important to get the relative ordering of the band head of the  $K=2^+$ band versus the ground and $K=0_2^+$ bands. 

\begin{figure*}[h]
\begin{minipage}{15pc}
\includegraphics[width=15pc]{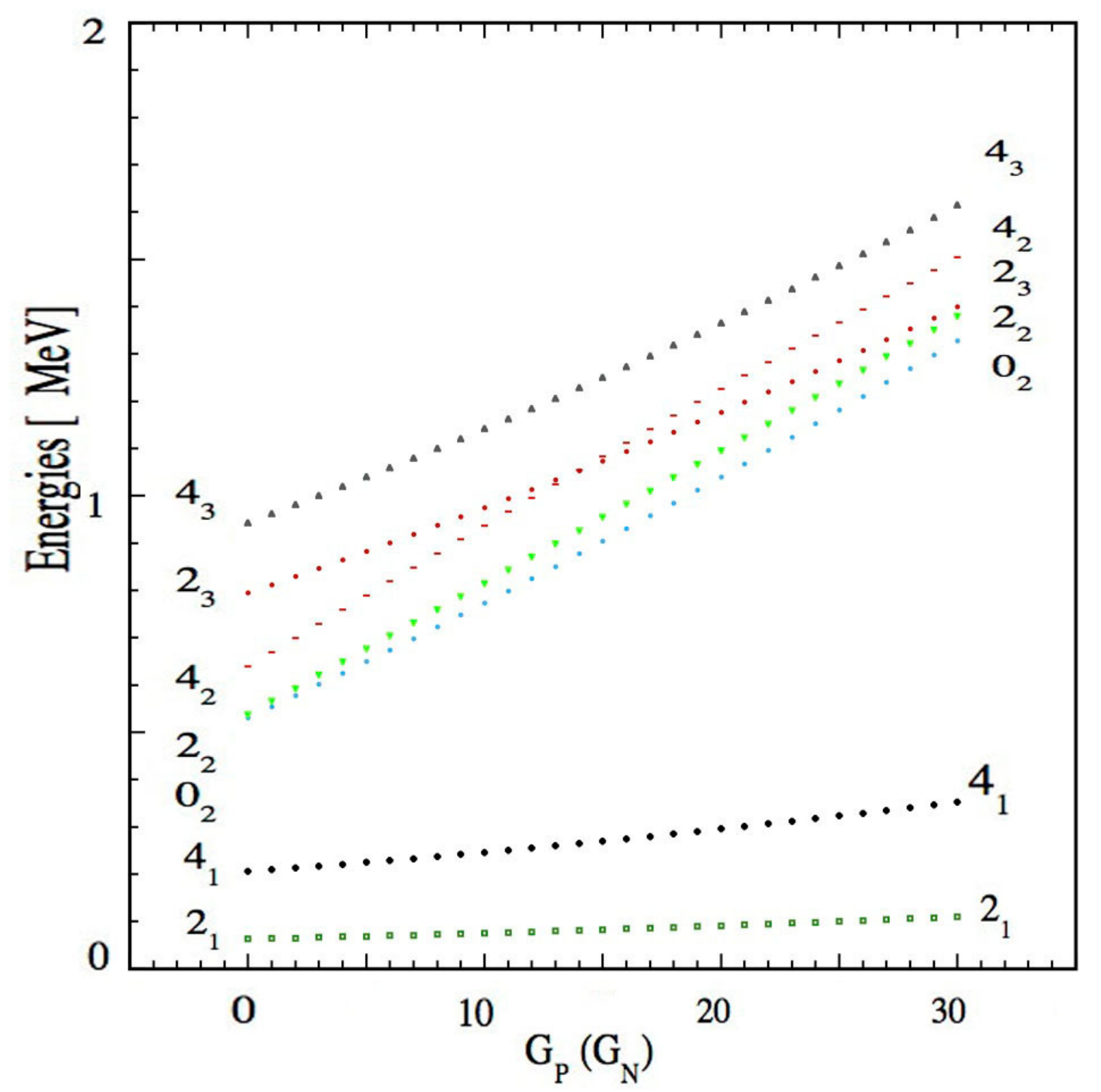}
\caption{\label{gd156envGp} The energy values (in MeV) of the few three states of total angular momentum 
$J = 0, 2, 4$,   versus the strength of the proton and neutron paring in $^{156}$Gd. }
\end{minipage}
\hspace{2pc}%
\begin{minipage}{20pc}
\includegraphics[width=20pc]{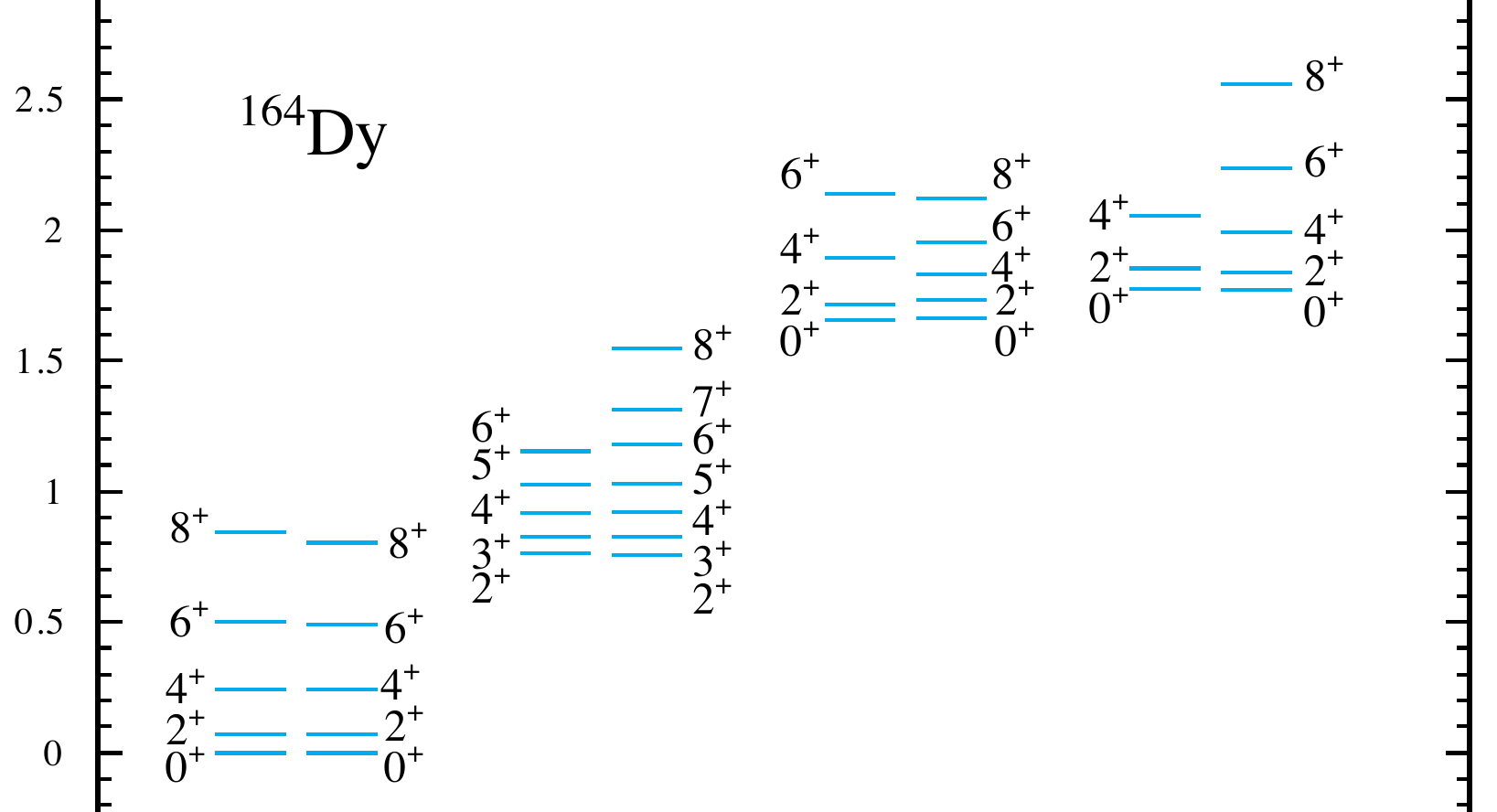}
\caption{\label{fig:dy}The energy values in MeV, in the  first four bands in $^{164}$Dy. The first one is the ground band, the second one is the $K=2^+$ band, the third and fourth are the $K=0_2^+$, and  $K=0_3^+$ bands. In the left hand side are given the experimental values\cite{TablesI}, and in the right the calculated ones.}
\end{minipage}
\end{figure*}

After varying one interaction strength at a time and studying the effect of each interaction on the
energy levels and B(E2) transitions, we
 fixed the values 
for pairing ($G_{\pi }=21/A$, $G_{\nu }=17/A$), as well as for the
quadrupole-quadrupole interaction strength ($\chi =35A^{-5/3}$)\cite{PHD00}.
The details of the study will be given elsewhere.
The  interaction strengths of the rotor terms were varied to give a best fit to
the the band heads of the first excited  $K^\pi=0^{+}$,
$K^\pi=2^{+}$ and $K^\pi=1^+$ bands, as well as the moment of
inertia of the g.s. band. Explicitly, the term
proportional to $K_J^2$ breaks the $SU(3)$ degeneracy of the
different K bands, the $J^2$ term represents a small correction to
fine tune the moment of inertia, and the last term, $C_2$, is
introduced to distinguish between $SU(3)$ irreps with $\lambda$
and $\mu$ both even from the others with one or both odd, hence
fine tuning the energy of the first excited $K^\pi=1^+$ state.

\begin{figure*}[h]
\begin{minipage}{18pc}
\includegraphics[width=18pc]{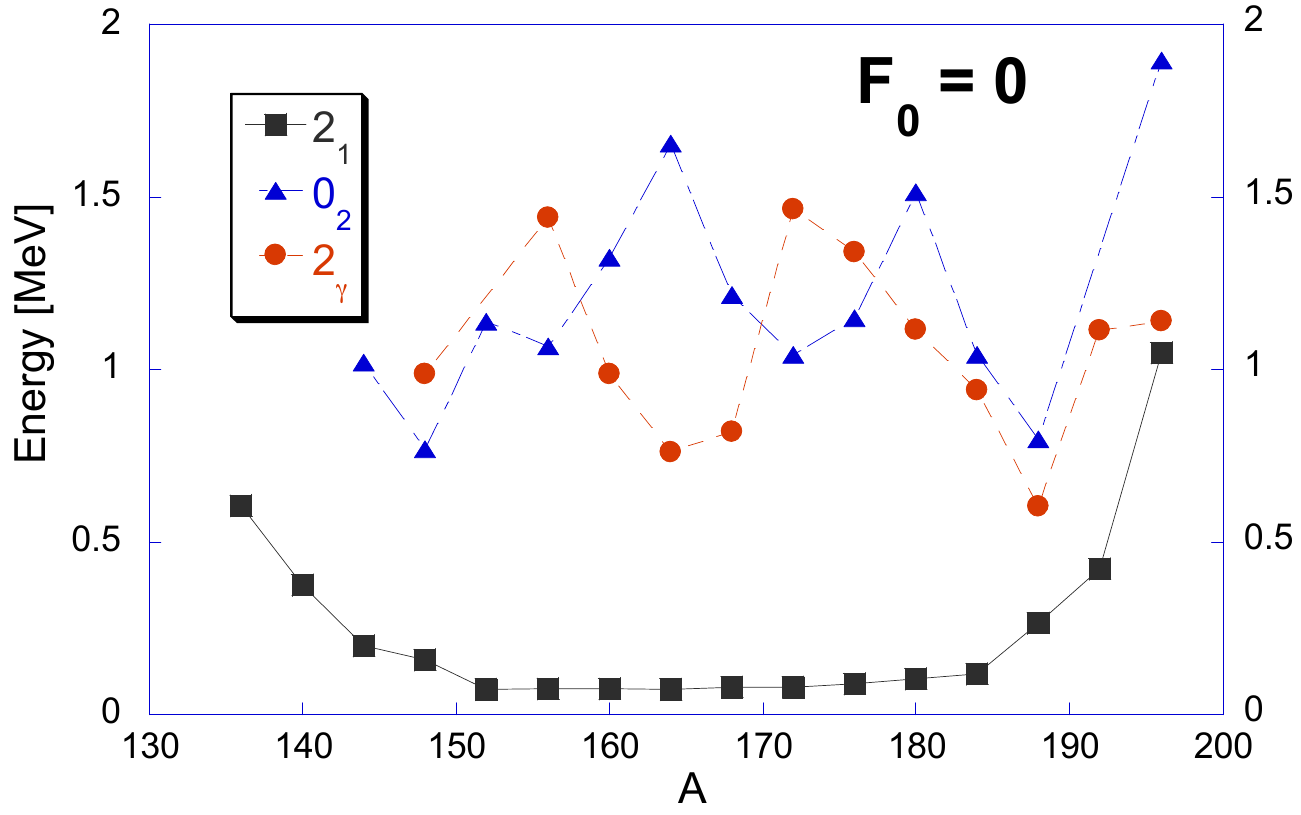}
\caption{\label{fig:1}The experimental energy values \cite{TablesI} of the
$J^\pi=2^+_1$, $J^\pi=0^+_{K^\pi=0_2^+}$, and
$J^\pi=2^+_{K=2^+_1}$ states in a sequence of nuclei for which the
numbers of valence protons and neutrons are the same, ($F_0=0$).}
\end{minipage}
\hspace{2pc}
\begin{minipage}{17pc}
\includegraphics[width=17pc]{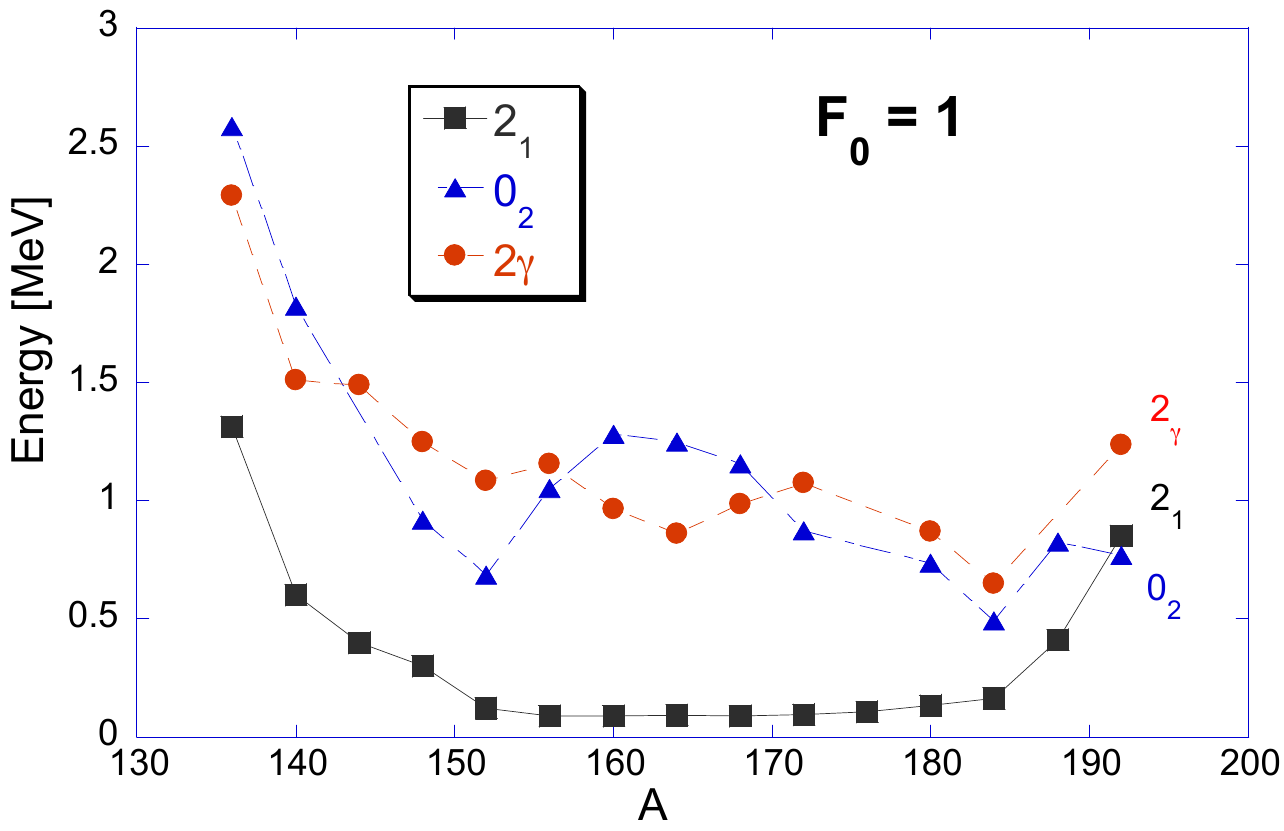}
\caption{\label{fig:2}The experimental energy values \cite{TablesI} of the
$J^\pi=2^+_1$, $J^\pi=0^+_{K^\pi=0_2^+}$, and
$J^\pi=2^+_{K=2^+_1}$ states in a sequence of nuclei for which the
difference in the numbers of valence proton and neutron pairs is
two, ($F_0=1$).}
\end{minipage}
\end{figure*}
With parameters fixed, the Hamiltonian is diagonalized. As an example, the energy values for the first four energy bands in $^{164}$Dy are given in Figure \ref{fig:dy}. In the  left hand side are given the experimental values\cite{TablesI}, and in the right the calculated ones.
Within this framework, the splitting and mixing of the
pseudo-$SU(3)$ irreps are generated by the proton and neutron
single particle terms ($H^{\pi/\nu}_{sp}$) and the pairing
interactions. This mixing plays an important role in the
reproduction of the behavior of the low-lying collective states in
deformed nuclei \cite{Popa04}. 

\begin{table*}[h]
\lineup
\caption{Energy values for the $J^\pi=2_1^+$, first excited
$J^\pi=0^+_{k^\pi=0^+_2}$ and $J^\pi=2^+_{K^\pi=2_1^+}$ states for
seven nuclei in the $F_0=0$ multiplet. The experimental values
\cite{TablesI} are given in parenthesis. The numbers of valence
proton and neutron pairs are given in the second and third
columns. The resulted fitted parameters of the rotor part of the Hamiltonian are given in the fourth to seventh columns.}
\label{tab:1}
\begin{center}
\begin{tabular}{llllllllll}
\br
nucleus & $N^\pi$ & $N^\nu$ &a & b& $a_s$& $a_3 $& $E(2_1)$ & $E(0^+_{K=0^+_2})$ & $E(2^+_{K^\pi=2^+_1})$  \\
&&&&&&$\times $& Th. (Exp.)& Th. (Exp.)& Th. (Exp.)\\
&&&&&&$10^{-4}$&[MeV]&[MeV]&[MeV]\\
\mr
$^{152}$Nd &5&5& 0.000& 0.00& 0.0000& 2.57   &0.082(0.073)&1.14(1.14)&1.31(1.38)\\
$^{156}$Sm &6&6& 0.000& 0.55& 0.0000& 2.59   &0.078(0.076)&1.07(1.07)&1.45(1.47)\\
$^{160}$Gd  &7&7& 0.001& 0.15 & 0.0034& 1.93  &0.085(0.075)&1.33(1.33)&0.99(0.82)\\
$^{164}$Dy  &8&8& -0.001& 0.04& 0.0008& 0.65  &0.073(0.073)&1.67(1.66)&0.76(0.76)\\
$^{168}$Er   &9&9& -0.002& 0.02& 0.0008& 0.75  &0.089(0.080)&1.21(1.22)&0.81(0.82)\\
$^{172}$Yb &10&10& -0.001& 0.12& 0.0010& 0.31&0.081(0.078)&1.04(1.04)&1.49(1.47)\\
$^{176}$Hf &11&11& -0.007& 0.30 & 0.0060& 0.43& 0.178(0.088)&1.15(1.15)&1.26\\
\br
\vspace{-6mm}
\end{tabular}
\end{center}
\end{table*}

\begin{table*}[h]
\caption{Same as in Table \ref{tab:1} for four nuclei with $F_0=1$. }
\begin{center}
\label{tab:2}
\begin{tabular}{llllllllll}
\br
nucleus & $N^\pi$ & $N^\nu$  &a & b& $a_s$& $a_3 $ & $E(2_1)$ & $E(0^+_{K=0^+_2})$ & $E(2^+_{K^\pi=2^+_1})$  \\
&&&&&&$\times $& Th. (Exp.)& Th. (Exp.)& Th. (Exp.)\\
&&&&&&$10^{-4}$& [MeV] & [MeV] & [MeV] \\
\mr
$^{156}$Gd & 7&5& 0.001 & 0.10 & 0.0014 & 1.36 & 0.094 (0.089)&1.06 (1.05)& 1.14 (1.15)\\
$^{160}$Dy &8&6 & 0.003 & 0.16 & 0.0004 & 0.94 & 0.092 (0.087)&1.30 (1.28)& 0.99 (0.97)\\
$^{168}$Yb &10&8& 0.00 & 0.01 & 0.0010 & 0.51 & 0.088 (0.088)&1.16 (1.15)& 0.98 (0.98)\\
$^{172}$Hf &11&9&  0.00 & 0.00 & 0.0000 & 0.51 & 0.119 (0.095)&0.87 (0.87)& 1.08 (1.07)\\
\br
\vspace{-10mm}
\end{tabular}
\end{center}
\end{table*}

This approach was employed for a systematic investigation of the experimental  energies of the
$J^\pi = 2_{g.s}^+$,  $J^\pi=0^+_{K^\pi=0_2^+}$,  and
$J^\pi=2^+_{K^\pi = 2^+_1}$ states in the deformed nuclei  from the $F_0=0$ and $F_0=1$  multiplets of the rare earths, 
 which are shown in Fig.\ref{fig:1} and Fig. \ref{fig:2}, where the rather irregular behavior of
the band-heads of the first excited bands is quite obvious.  The calculated energies 
in the framework of pseudo-SU(3)
model, with above described fit of the respective parameter strengths,  
are compared to the experimental values in Tables
\ref{tab:1} and \ref{tab:2}.  We calculated also all the states
with $J \le 8$ within these three bands.
The obtained results are in very good agreement with experiment
\cite{Popa04,TablesI}. The main reason for obtaining the position
of each collective band with respect to each other, as well as of
each level within the band is the specific content of the obtained
SU(3) irreps into the collective states, which is related to their
deformations. For nuclei from Table \ref{tab:1}, in the middle of
the shell, the ground  and the $\gamma$ band belong to the same
($\lambda ,\mu$) with $\lambda > \mu$. At the limits of the
deformed  region the ground band states have oblate deformation
($\lambda < \mu$) and the $K^\pi = 0_2^+$ and the $K^\pi = 2^+_1$
bands are mixed in the same SU(3) irrep \cite{Popa04}. The
analysis for nuclei from Table \ref{tab:2} is under investigation.

The correct description of collective properties of  first excited
$K^\pi=0_2^+$,  and $K^\pi=2^+$  states is  a result of
representation  mixing and state deformation induced by the model
Hamiltonian. This study shows that pseudo-spin zero neutron and
proton configurations with relatively few pseudo-SU(3) irreps with
largest C2 values suffices to yield good agreement with known
experimental energies.

In the presented applications of the pseudo-SU(3) model, we achieved a good description of the complex 
behavior of a series of heavy deformed nuclei using a small configuration space. We are working on further analyzing the 
effect of the different components of the interaction in the Hamiltonian, on the strength of the transition probabilities.
Also, the size of the model space will be further investigated. 
Work in progress is on expanding the study to other groups of nuclei  related to one another by an over-arching F-type symmetry.

\ack{}
This work is an appreciation of the numerous contributions of J. P. Draayer  and his
collaborators to the deeper understanding from symmetry principles of the nuclear 
structure, and the inspiration he provided in this way to the authors. We also acknowledge
the support from the Bulgarian National
Foundation (DID-02/16) and the Bulgarian-Rumanian agreement (DNTS-02/21),
the U.S. National Science Foundation (PHY-0904874), Department of Energy (DE-SC0005248), and the
Southeastern Universities Research Association.


\end{document}